\documentstyle[11pt,newpasp,twoside,epsf]{article}
\def\edcomment#1{\iffalse\marginpar{\raggedright\sl#1\/}\else\relax\fi}
\reversemarginpar

\begin{document}
\title{An All-sky Advanced X-ray Monitor (AXM) Mission}
 \author{Hale Bradt}
\affil{Massachusetts Institute of Technology, Room 37-587, Cambridge MA 02139-4307, USA}
\author{Ronald A. Remillard}
\affil{Massachusetts Institute of Technology, Room 37-595, Cambridge MA 02139-4307, USA}

\begin{abstract}
A concept for an all-sky monitor mission substantially more powerful than the RXTE ASM is presented in brief. It makes use of 31 Gas Electron Multiplier (GEM) proportional counters to provide continuous monitoring of the entire x-ray sky, except for earth occultations and a small region around the sun. This coverage would allow the study of a wide variety of short duration cosmic outbursts. At time scales longer than a few minutes, the sensitivity for the monitoring of x-ray sources (1.5 -- 15 keV) would be greatly improved over the RXTE ASM, a factor of 6 in one day. This mission concept, the Advanced X-ray Monitor (AXM), is described in more detail in Remillard et al. (2000).
\end{abstract}

\section{Introduction}
The All-Sky Monitor (ASM) on the RXTE (Levine et al. 1996) has proven remarkedly successful in (1) guiding observers to the optimum use of pointed instruments, both ground based and spaceborne, (2) providing context for short pointed observations, (3) revealing periodicities and aperiodic behavior on time scales from hours to years, and (4) the detection of, and monitoring of, explosive behavior in the universe. The latter include outbursts in galactic microquasars, the location and x-ray light curves of gamma-ray bursts, and outbursts in BL Lac objects. The RXTE ASM was launched in Dec. 1995 and is still operative with only modest degradation of its performance. Nevertheless, it is appropriate to contemplate the design of a follow-on experiment which we call the Advanced X-ray Monitor (AXM).

\section{The AXM mission}
The study of long term variability (items 1-3 above) can be carried out with a variety of approaches. However, the study of short-duration explosive phenomena (item 4 above) places significant constraints on the design. Rare events occurring at unexpected locations on the celestial sphere are best captured with an instrument that views continuously a large part of the sky. The optical, radio or x-ray identification of such events often requires a positioning capability of a few arcmin or less. There are also temporal variations associated with cosmic explosions that can help reveal the nature of the explosive sources. A coded mask can provide both the needed angular precision and the needed instantaneous large effective area and is thus preferable to the single pin-hole approach. The desire to study short explosive events, and the features required to attain it, distinguish the AXM concept from those of the forthcoming MAXI experiment on the International Space Station or the long awaited MOXE experiment on Spectrum X/gamma.

Our concept therefore builds upon the ASM design by incorporating 31 individual x-ray cameras on a Small Explorer (SMEX) mission (Figure 1), each with effective area about 2.3 times that of one of the three RXTE ASM cameras. Each camera (Figure 2) consists of a two-dimensional imaging proportional counter with a two dimensional random mask with 12 arcmin mask elements and a total field of view of 40\deg (FWHM). 

\begin{figure}
\plotone{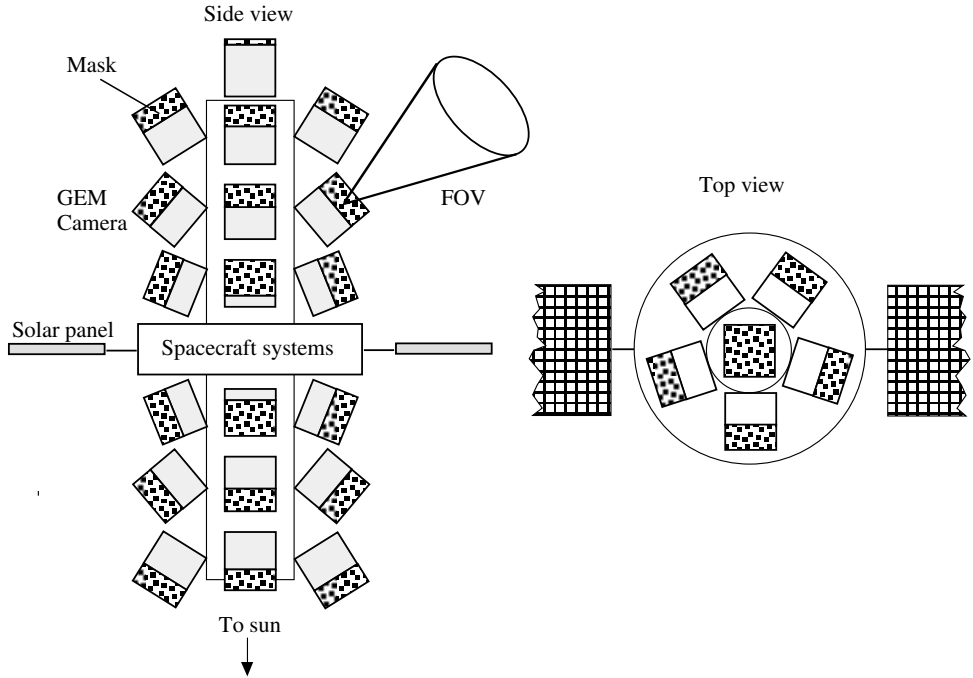}
\caption{The AXM spacecraft. The side view shows the 19 of the 31 cameras that view, at least in part, the hemisphere of the reader. The top view shows the topmost layer of cameras.}
\end{figure}

\begin{figure}
\plotone{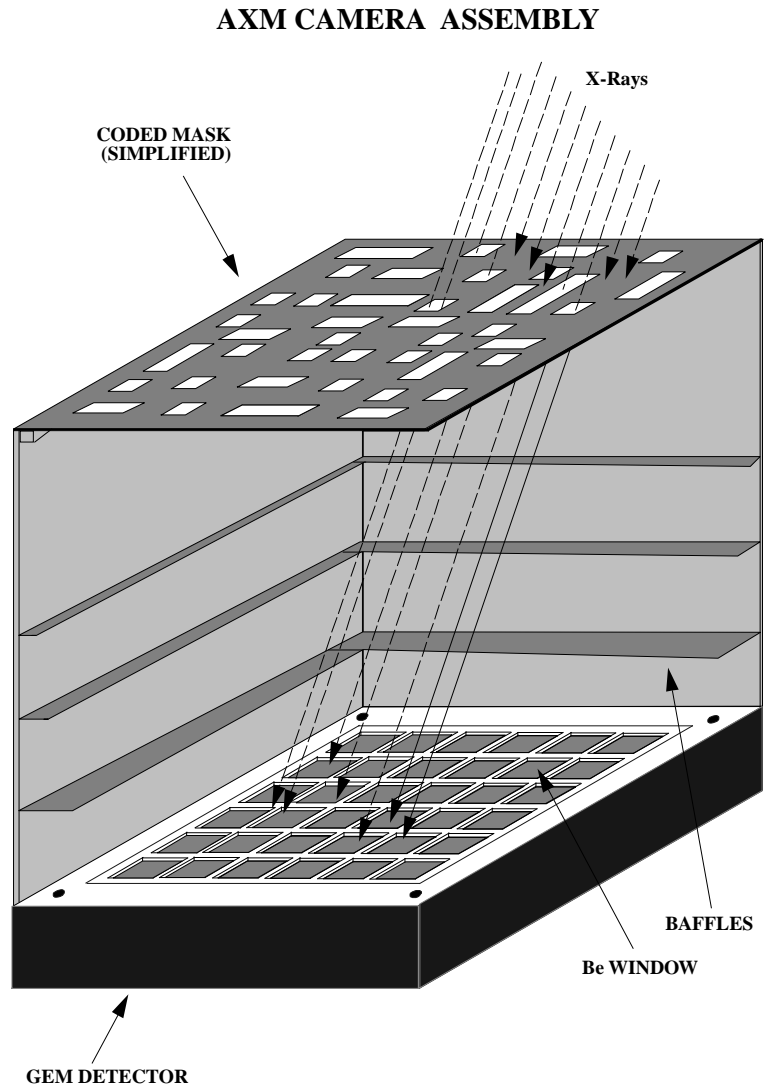}
\caption{One of the 31 GEM cameras}
\end{figure}

The pointing directions of the 31 cameras are distributed uniformly, such that, taken together, the fields of view encompass cover 97\% of 4$\pi$ steradians, the missing 3\% being located at the sun. In a low earth orbit, the earth blockage reduces the instantaneous sky coverage to 64\%. The positioning capability for a bright source is about 1 arcmin. The 5 $\sigma$ sensitivity to a very brief 1--s flare in a previously unknown source is 425 mCrab, or 300 mCrab for a known source, while it is only 0.8 mCrab (3 $\sigma$) for a persistent source viewed for one day.

The large number of detectors requires a robust design. The development of two-dimensional detectors for space missions has consistently been fraught with difficulties, with only a few detectors being flown on any given mission. We believe that Gas Electron Multiplier (GEM) proportional counters, recently developed and coming into use by the high energy physics community (Sauli 1997, Bachmann et al. 1999), are a promising answer to this problem. These proportional counters provide electron multiplication within the many closely-spaced small holes in one or two thin membranes stretched across the interior of the counter (Figure 3). Each membrane is a thin plastic coated on each side with a thin conducting metal. The counter geometry naturally maintains the x,y coordinate of the original energy deposit, while the spacing between the holes, $\sim$0.1 mm, yields good spatial resolution. This scheme serves to separate the multiplication regions from the charge collection grids, thus eliminating the risk of breakdown at the collection points.

\begin{figure}
\plotone{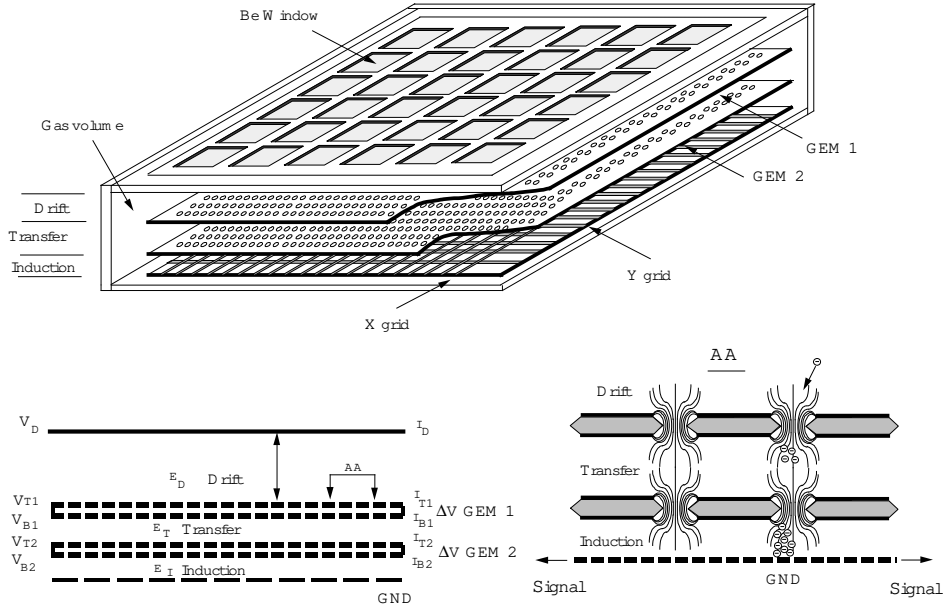}
\caption{Schematic of detector with two GEM multiplications stages.}
\end{figure}

The science objectives of the mission fall into three categories: 

(1) The science of jets and other cosmic explosions. These include ejections of relativistic jets in microquasars, flashes in fast x-ray novae, flashes from soft gamma-ray repeaters (SGRs), and gamma-ray bursts (GRB). These various types of events can take place on timescales from second to hours and can occur at unexpected times and locations in the sky. This has greatly impeded their study. Figure 4 shows simulations of three different types of actual outbursts, as they would appear in AXM data.

\begin{figure}
\plotone{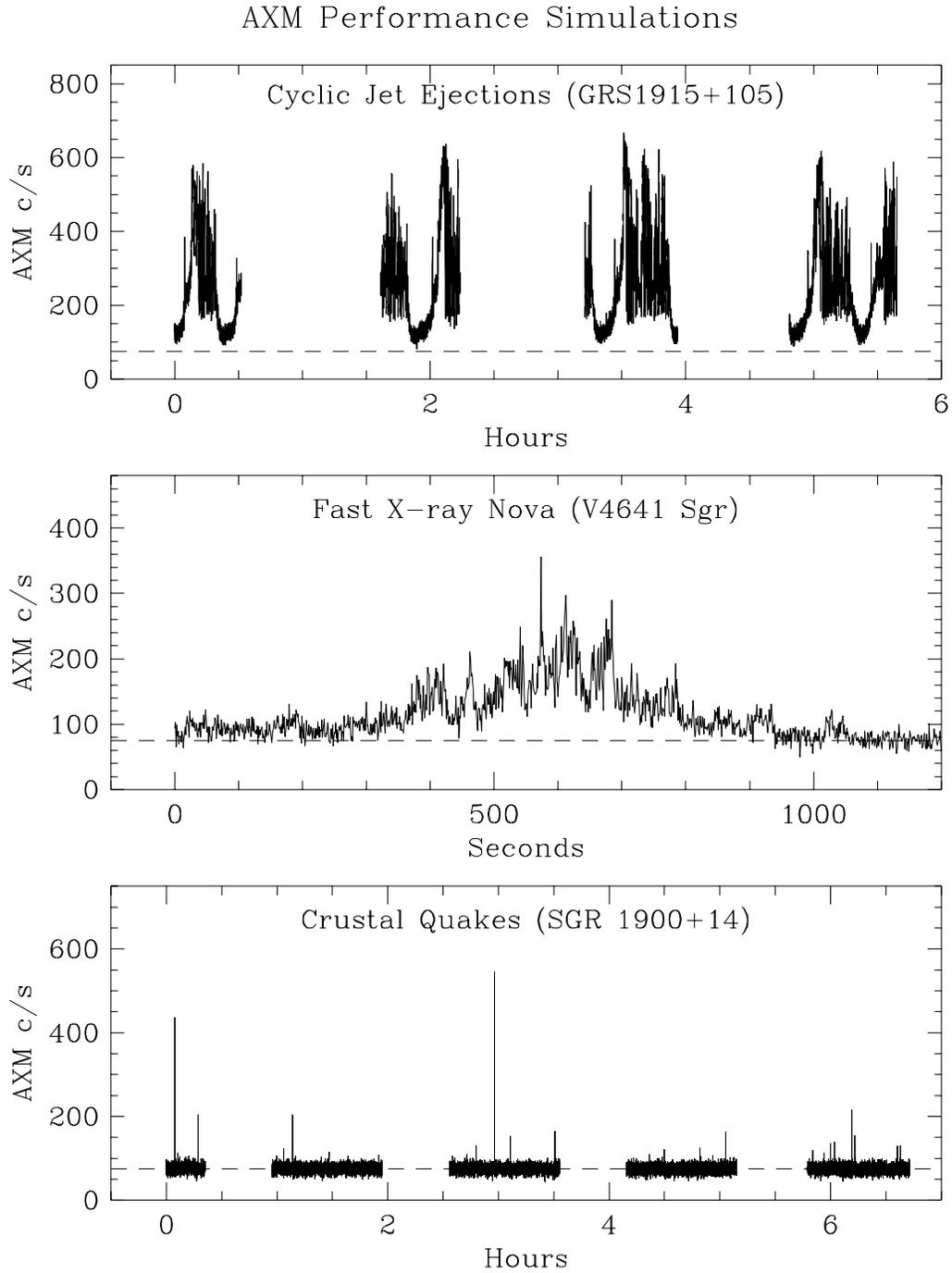}
\caption{Simulations with counting statistics of AXM performance. The simulations are based on actual events detected with the RXTE PCA.}
\end{figure}

(2) Long term spectral evolution of persistent sources. The AXM would produce light curves that show variability in beautiful detail, e.g., tracing the spectral evolution of black hole binaries or tracking excursions of accretion flow in low-magnetic-field neutron star systems. Spin changes would be monitored for accretion-powered pulsars to test evolving models of accretion torques and disk-magnetosphere interaction. Pulsar spin changes in the anomalous x-ray pulsars (AXP) would test the magnetar hypothesis and the possible association with SGRs. The mission would also play a major role in the study of variability in active galactic nuclei (AGN), in particular TeV outbursts in BL Lac objects. The associated x-ray outbursts would  probe outburst mechanisms and would guide the observing programs of TeV observatories, thus greatly increasing their efficiencies.

(3) Empowerment of other observatories and multifrequency science. The continuous knowledge of the state of the x-ray sky can guide and enhance the efficiency of missions such as Chandra, XMM, Integral, and GLAST, as well as observatories at optical and radio wavelengths. The x-ray light curves will provide context for brief pointed observations at all wavelengths.

\section{Summary}
The mission described here would serve as a major support to other missions in the coming decade but would also be a primary tool for the study of explosive phenomena in the universe. Its realization requires the development of GEM detectors qualified for space flight. Many persons have contributed to this concept for an AXM; the principle players are the authors of Remillard et al. (2000).

\end{document}